# Coulomb pairing resonances in multiple-ring aromatic molecules

D.L. Huber*

Physics Department, University of Wisconsin-Madison, Madison, Wisconsin 53706, USA

**Abstract**

We present an analysis of pairing resonances observed in photo-double-ionization studies of $C_nH_m$ aromatic molecules with multiple benzene-like rings. The analysis, which is based on the Coulomb pairing model, is applied to naphthalene, anthracene, phenanthrene, pyrene and coronene, all of which have six-member rings, and azulene which is comprised of a five-member and a seven-member ring. There is a high energy resonance at ~ 40 eV that is found in all of the molecules cited and is associated with paired electrons localized on carbon sites on the perimeter of the molecule, each of which having two carbon sites as nearest neighbors. The low energy resonance at 10 eV, which is found only in pyrene and coronene, is attributed to the formation of paired HOMO electrons localized on arrays of interior carbon atoms that have the point symmetry of the molecule with each carbon atom having three nearest neighbors. The origin of the anomalous increase in the doubly charged to singly charged parent-ion ratio that is found above the 40 eV resonance in all of the cited molecules except coronene is discussed.

*Mailing address: Physics Department, University of Wisconsin-Madison, 1150 University Ave., Madison, WI 53711, USA; e-mail: dhuber@wisc.edu

## 1. Introduction

Recent studies of photo-double-ionization in $C_nH_m$ multiple-ring (polycylic) aromatic molecules have revealed the existence of anomalous resonances in the ratio of the cross sections of doubly charged parent ions to singly charged parent ions $I(2+)/I(1+)$ [1-4]. In the majority of the molecules where the resonances have been found, only a single peak is observed at ~ 40 eV above threshold. The notable exceptions are coronene and pyrene [1,3,4]. In these compounds, there is an additional resonance at 10 eV above threshold. Two explanations have been proposed for the origin of the resonance: Cooper pairing, the mechanism for superconductivity [5], and Coulomb pairing [6]. In this note we use the Coulomb pairing model in analyzing the origin of both the high energy and low energy resonances and how they relate to the geometry of the molecule.

As discussed in [6], Coulomb pairing was discovered in a study of the two-electron excitations in a one dimensional ‒metal‖ with periodic boundary conditions [7,8]. In applying the theory to planar molecules, the corresponding boundary condition is $C_n$ point symmetry about an axis perpendicular to the plane [9]. In addition, it is necessary that the holes that are created in the HOMO band when the two electrons are excited are delocalized so as to rule out Frenkel exciton-like excitations where the hole and the excited electron are on the same or neighboring carbon sites. When these conditions are satisfied, there can be paired spin singlet states with energies on the order of several 10s of eV above the LUMO band and characteristic lengths on the scale of the C-C separation [6]. In [1,2] the authors point out that the energy of the 40 eV resonance is related to the distance between neighboring carbon atoms. To see the connection, we make the assumption that the Coulomb pair decays into a zero total momentum state with the two free electrons having the same kinetic energy. We express the energy in terms of the electron mass and the de Broglie wavelength, $\lambda_{DB}$. With a resonance energy of 40 eV, we find $\lambda_{DB} = 2.6$ Å, a value that is approximately twice the C-C separation of 1.4 Å. A similar argument can be made for the 10 eV resonance, with result that $\lambda_{DB}$ in this case is 5.2 Å. Expanding the localized state wave function, characterized by the length scale $\lambda_{LOC}$, in terms of the symmetrized two-electron plane-wave functions, it becomes evident that plane-wave states with $\lambda_{DB} << \lambda_{LOC}$ do not make a significant contribution to the expansion. It follows that the maximum spatial extent of the pair is on the order of $\lambda_{DB}$. It should be noted that the introduction of the de Broglie wave presented here pertains to the ‒free‖ electrons that result from the break-up of the bound pair. In contrast, in [1,2] the de Broglie wavelength is associated with the kinetic energy of the bound pair.

## 2. Analysis

In analyzing the photo-double-ionization results, the focus is on the parameter $R$ defined by

$$R = [I(2+)/I(1+) - "knockout"] \qquad (1)$$

where *"knockout"* is the scaled double-to-single photoionization ratio of helium [4]. When the values of $R$ for naphthalene, anthracene, phenanthrene, and azulene are plotted as a function of the excess energy, all of the molecules display a peak at ~ 40 eV [1-4] which is close to the position of the peak in deuterated benzene [2,4]. Of polycyclic molecules, naphthalene and anthracene have $C_2$ point symmetry, whereas phenanthrene and azulene lack rotational symmetry altogether. From the molecular diagrams displayed in Fig. 1, it is evident that all of the molecules except benzene have a multiple ring structure made up of segments of carbon (C) sites, each with two nearest-neighbors that are bounded by C sites with three nearest neighbors. We label the carbon sites with two nearest neighbors C(2) sites, and the sites with three nearest neighbors C(3) sites. In these terms, benzene has six C(2) sites and no C(3) sites whereas naphthalene has eight C(2) sites and two C(3) sites.

In [6] it was pointed out that the paired state wave function in benzene is strongly localized, with an amplitude that decreases exponentially with the number of intervening bonds. The fundamental hypothesis in the multiple-ring theory is that the high energy resonance is associated with localized states of the paired electrons that are confined to regions where there are two or more connected C(2) sites. The paired electrons occupy segments of a benzene-like ring that are bounded by C(3) sites. In naphthalene there are two segments of four sites that meet this condition. In anthracene, the two end rings have the same properties as the rings in naphthalene while the middle ring has no sites that meet this condition. In the case of phenanthrene, the end rings are the same as in naphthalene, while the middle ring has a segment with two C(2) sites. Azulene has one segment with three connected C(2) sites and one segment with five connected C(2) sites.

As mentioned, pyrene and coronene have both a high energy and a low energy resonance. The perimeter of the pyrene molecule shown in Fig. 2 has two segments with two connected C(2) sites and two segments with three connected C(2) sites. In the case of coronene, the perimeter has six segments with two connected C(2) sites, In both molecules, the perimeter segments are bounded by C(3) sites. The similarity between the perimeter segments in pyrene and coronene and the segments in naphthalene, anthracene, phenanthrene, and azulene supports the argument for a common origin of the high energy resonance.

**3. Pyrene and coronene**

In the case of pyrene, we argue that the low energy pairing is associated with the two inner C(3) sites that are connected by the $C_2$ point symmetry transformation. In coronene, the pairing is associated with the six C(3) sites in the interior that are connected by the $C_6$ point symmetry transformation. None of the molecules that display only a 40 eV resonance have similar internal configurations. Also, with three nearest neighbors, the wave function of the Coulomb pair will be expanded relative to the wave function in the two-nearest-neighbor configuration which will reduce the effects of the electrostatic interaction between the pairing electrons and thus lower the energy of the resonance. Taken together, these two points strengthen the argument for associating the low energy resonance with electron pairs localized inside the perimeters of the molecules.

**4. Discussion**

The preceding analysis leads to a qualitative picture of the photo-double-ionization resonances in multiple-ring aromatic molecules that is based on the Coulomb pairing of the HOMO electrons. As shown previously [6,8], the electronic wave function of the pair is strongly localized. An essential assumption in the analysis is that there are no paired states involving a C(3) site and a C(2) site. Since the C(2)-C(3)-C(2) configuration is limited to the perimeter of the molecule (Figs. 1 and 2), the assumption has the effect of dividing the perimeter into benzene-like segments with one or more C(2) sites. Because of the strong localization, the energies of the resonances in the segments with two or more C(2) sites are close to the energy of the resonance in benzene with its ring of six C(2) sites [1]. Although the assumption is that there is no pairing involving a C(3) site that has two C(2) sites as nearest neighbors, there can be pairing involving interior nearest-neighbor C(3) sites as occurs in pyrene and on sites on the inner benzene-like C(3) ring in coronene (Fig. 2).

As mentioned, phenanthrene and azulene lack $C_n$ symmetry but still show resonances at 40 eV. We attribute the anomalous behavior to the fact both molecules have end rings with 4 (phenanthrene) or 7 and 3 (azulene) connected C(2) sites. The C(3) sites that bound the segments can be viewed as perturbations of the $C_n$ symmetry that are too weak to destroy the pairing since the maximum spatial extent of the pair is on the order of the de Broglie wave length which is about twice the C-C separation.

It is pointed out in [164] that all of the molecules cited previously, with the exception of coronene, show an anomalous increase in the parameter *R* starting at the 40 eV resonance. Although we don¢t have a detailed explanation for this increase, we note that coronene is the one molecule among those cited where there is only the configuration C(3)-C(2)-C(2)-C(3) in the perimeter segments. The other molecules all have at least one perimeter segment with more than two connected C(2) sites. We conjecture that the anomalous increase in *R* is a many-body effect arising from the presence of more than two connected C(2) sites in a single perimeter segment.

As a final comment, we note our analysis leads to the important prediction that all $C_nH_m$ multiple-ring aromatic molecules will have a photo-double-ionization resonance at ~ 40 eV above threshold. It also provides an explanation for the 10 eV resonances seen in pyrene and coronene and points to a possible explanation for the absence of an increase in *R* above 40 eV in the latter molecule.

**Acknowledgement**

We thank R. Wehlitz for helpful comments on the interpretation of the photo-double-ionization measurements.

**Figure captions**

Fig. 1. Benzene, naphthalene, anthracene, phenanthrene and azulene. The symbols C(2) and C(3) characterized a segment of the naphthalene perimeter of the form C(3)-C(2)-C(2)-C(2)-C(2)-C(3), where C(n) denotes a carbon site with n nearest neighbors.

Fig. 2. Pyrene: there are six C(3) sites. The low energy pairing involves the two inner C(3) sites connected by the "**P**" bond. Coronene: There are twelve C(3) sites. "**P** ring" refers to the inner low energy pairing ring of six C(3) sites.

**Benzene**

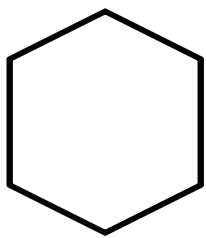

**Naphthalene**

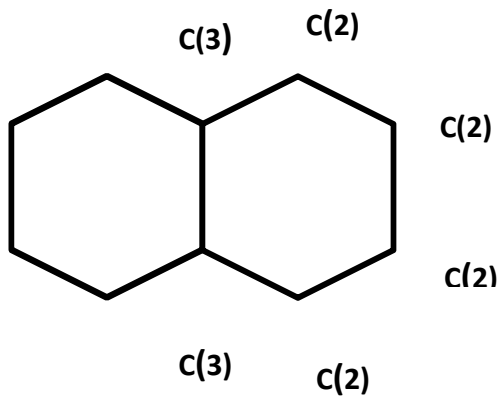

**Anthracene**

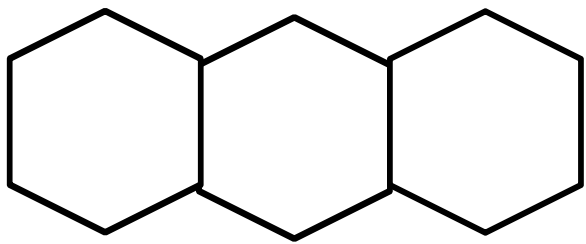

**Phenanthrene**                                                **Azulene**

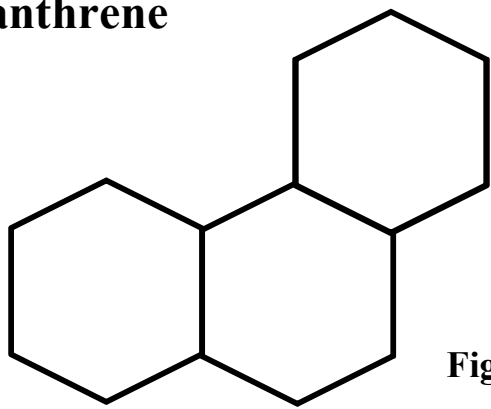              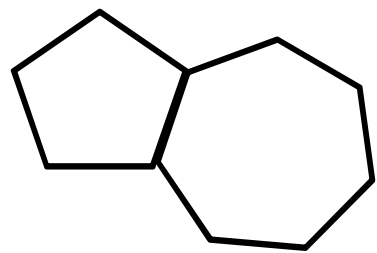

Fig. 1

## Pyrene

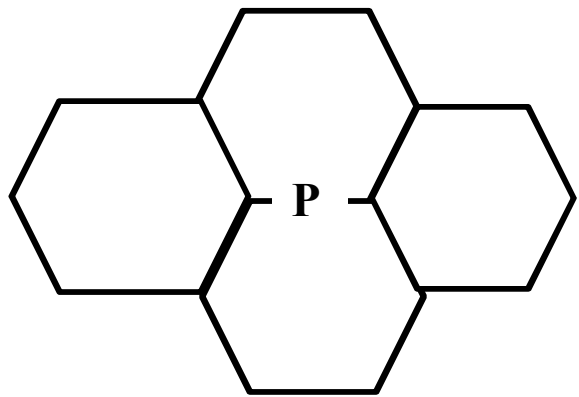

## Coronene

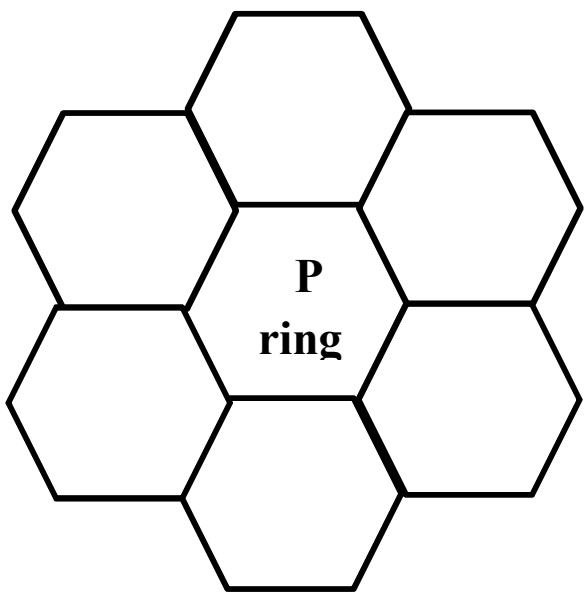

**Fig. 2**